\documentclass[11pt,preprint,graphicx]{aastex63}

\usepackage[fleqn]{amsmath} 
\usepackage{amsfonts} 
\usepackage{amstext}
\usepackage{amssymb} 
\usepackage{natbib}
\usepackage{graphicx}
\usepackage{lineno}

\newcommand{\hi } {{\rm H}\,{\small\rm I} \,}
\def\etal{\it et al. \rm }

\begin{document}

\title{The Baryonic Tully-Fisher Relation II: Stellar Mass Models}

\author{Francis Duey}
\affiliation{Department of Astronomy, Case Western Reserve University, Cleveland, OH 44106}
\author[0000-0003-2022-1911]{James Schombert}
\affiliation{Dept. of Physics, University of Oregon, Eugene, OR 97403}
\author[0000-0002-9762-0980]{Stacy McGaugh}
\affiliation{Department of Astronomy, Case Western Reserve University, Cleveland, OH 44106}
\author[0000-0002-9024-9883]{Federico Lelli}
\affiliation{Arcetri Astrophysical Observatory (INAF), Florence, Tuscany, IT}

\begin{abstract}

\noindent We present new color-$\Upsilon_*$ (mass-to-light) models to convert WISE W1
fluxes into stellar masses.  We outline a range of possible star formation histories
and chemical evolution scenarios to explore the confidence limits of stellar
population models on the value of $\Upsilon_*$.  We conclude that the greatest
uncertainties (around 0.1 dex in $\Upsilon_*$) occur for the bluest galaxies with the
strongest variation in recent star formation.  For high mass galaxies, the greatest
uncertainty arises from the proper treatment of bulge/disk separation in which to
apply different $\Upsilon_*$ relations appropriate for those differing underlying
stellar populations.  We compare our deduced stellar masses with those deduced from
{\it Spitzer} 3.6$\mu$m fluxes and stellar mass estimates in the literature using
optical photometry and different $\Upsilon_*$ modeling.  We find the correspondence
to be excellent, arguing that rest-frame near-IR photometry is still more
advantageous than other wavelengths.

\end{abstract}

\section{Introduction}

The conversion of stellar luminosity into stellar mass is the keystone to
understanding the total baryonic mass of a galaxy (Tinsley 1968) and its mass
distribution (Ibarra-Medel \etal 2015).  In addition, the stellar mass of a galaxy is
the primary output of its star formation history, the conversion of neutral gas into
stars, and the defining characteristic for the evolution of a galaxy (Speagle \etal
2014).  The production, or lack of production, of stars defines a galaxy's morphology
(Cassata \etal 2007), color evolution (Rakos \& Schombert 1995) and metallicity
evolution (Baker \& Maiolino 2023).  Whether the growth of stellar mass happens in a
monolithic (Eggen, Lynden-Bell \& Sandage 1962) or hierarchical fashion (Treu \etal
2005), our understanding of the baryon cycle begins with a galaxy's current stellar
mass.

The path to determining stellar mass begins with photometry, as the luminosity of a
galaxy is simply a proxy for the number of stars.  The photometry of galaxies has a
long history (Roth 2023) stretching from uncalibrated photographic imaging through
the era of photoelectric photometry that produced catalogs such as the RC3 (de
Vaucouleurs \etal 1991) to the present-day digital sky surveys from the ground (e.g.,
SDSS, Consolandi \etal 2016) and space (e.g., WISE, Wright \etal 2010).  However,
even for an isolated galaxy with few nearby stellar sources, the determination of an
accurate flux (total or by area, i.e. surface photometry) is an observational
challenge (Hogg 2022).  An additional dilemma arises with any attempt to extract a
true bolometric flux from photometric magnitudes without some knowledge of the total
energy flux emitted across all wavelengths (Brown \etal 2014).

As noted by Taylor \etal (2011), almost all the characteristics of a galaxy,
including its morphological appearance, vary sharply during the course of its
evolution.  Detailed analysis of the color-magnitude diagrams of stellar populations
in nearby galaxies (Weisz \etal 2011) shows that the star formation history of a
galaxy may change dramatically, often on short timescales, along with its color and
mean parameters such as stellar population age and metallicity.  The global
properties of galaxies, including their kinematics, are well correlated with their
total stellar masses, regardless of their suspected paths of galaxy evolution (see
Strateva \etal 2001 and Gallazzi \etal 2006).  The exploration of these stellar mass
scaling relations back to the era of galaxy formation is a primary science goal of
the recently launched James Webb Space Telescope (Dressler \etal 2023).

The process of converting galaxy fluxes into stellar mass requires interpretation
with a stellar population model.  These models take a unit amount of gas mass and an
initial mass function, plus an assumed star formation history and chemical enrichment
scenario, and combine these functions with known stellar isochrones to produce a mean
mass-to-light ratio (hereafter $\Upsilon_*$).  The effects of age and metallicity on
the $\Upsilon_*$ of stellar populations are obvious from even the simplest star
formation scenarios.  For example, a young stellar population rich in high mass
O-stars will have a particularly low $\Upsilon_*$ due to the enormous luminosity of
O-stars as compared to an old stellar population dominated by lower luminosity red
giant branch stars, which will drive $\Upsilon_*$ to higher values (Tinsley 1978).  

At the core of these population models are the simple stellar populations (SSP),
stellar groups of single age and metallicity (Bruzual 1993).  While the colors and
spectral energy distributions (SED) of SSP's are well matched to the observations of
compact systems, such as globular clusters, they fail to adequately match the colors
of normal galaxies due to the complex mixing of many populations of different ages
and metallicities (Tinsley 1980).  A real, present-day galaxy will have a stellar
population that evolved over time in two basic ways, 1) a changing star formation
rate (the star formation history, SFH) plus 2) a growing mean metallicity as previous
generations have enriched the ISM from which the next generation of stars form
(chemical evolution).  Basic composite stellar populations (CSP), ones which are
linear combinations of SSP's, produce high quality matches to the color and spectroscopic
characteristics of normal galaxies (Bell \etal 2003, Gallazzi \etal 2005) and provide
a testable pathway to mapping $\Upsilon_*$ onto observables such as morphological
type or mean color (Walcher \etal 2011, Schombert, McGaugh \& Lelli 2022).

An additional level of accuracy is obtained when CSP models are matched to the galaxy
observables such as a full SED, various spectral indices or widely spaced colors to
limit the range of properties of the underlying stars.  Each technique has its
various strengths and inherent uncertainties.  For example, one of the finest
analysis comparing synthetic SED's to multiband photometry is the GAMA project
(Driver \etal 2009) outlined in detail in Taylor \etal (2011).  Their technique of
fitting the flux in nine bandpasses from the UV to the near-IR produces uncertainties
in stellar mass of only 0.12 dex (28\%).  While they conclude that the addition of
mid-IR fluxes adds additional uncertainty, probably due to the luminosity changes
induced by short-lived asymptotic giant branch stars (AGB), they found acceptable
errors when just using SDSS fluxes and a $g-i$ color relation to deduce a
$\Upsilon_*$.

This is the second paper our series to study of one of the key galaxy scaling
relations, the baryonic Tully-Fisher relation.  We have dedicated this paper to the
construction of an observational technique for the conversion from flux densities
(centered on the WISE mission photometric archives) to stellar mass which, when
combined with the gas mass, becomes the total baryonic mass of a galaxy.  Our first
paper in this series (Duey \etal 2024) outlined our procedures and protocols for
using WISE and {\it Spitzer} imaging to assign a total galaxy luminosity.  The goal
of this paper is to 1) introduce new mass-to-light models for the WISE filters, 2)
compare these models to {\it Spitzer} and SDSS stellar masses and 3) map the
resulting luminosity TF relation into the stellar mass TF, popular in distance scale
studies (Tully \etal 2009).  The third paper in this series will be to take the
baryon masses deduced in Papers 1 and 2, combined with redshift independent
distances, to definitively characterize the IR baryonic Tully-Fisher relation in
order to explore the kinematic scaling relations of disk galaxies and standardize the
baryonic TF for distance scale work.

\section{WISE Mass-to-Light Models}

For stellar mass determination, the near-IR offers a portion of a galaxy's spectrum
that is dominated by starlight free from strong emission lines and with limited
changes due to young stars.  In addition, the low Galactic and internal extinction
corrections make near-IR fluxes more consistent across morphological types.  The
SPARC project depended on pointed observations from the {\it Spitzer} Space Telescope
(Werner \etal 2004). With the termination of the {\it Spitzer} mission, future
dynamical studies from our team will use the all-sky dataset from the WISE
(Wide-field Infrared Survey Explorer) mission (Wright \etal 2010).  Fortunately,
the WISE and {\it Spitzer} filters overlap in the near-IR.

The last step to converting a luminosity to a stellar mass is an accurate
mass-to-light ratio, $\Upsilon_*$.  The mass and luminosity of a single
star is a straight forward calculation deduced from the field of stellar structure
and has been calibrated with binary stars.  The determination of the mass of a star
cluster is also a well solved problem if the age and metallicity of the stellar
population is known (by, for example, matching isochrones to the cluster
color-magnitude diagram).  The integrated mass of a SSP
can be deduced from the application of computed stellar evolutionary tracks and an
assumed initial mass function, both of which can be tested by detailed
color-magnitude diagrams (see Hosek \etal 2020) and various secondary spectral
indices (see Trager \etal 1998).  The procedure becomes much more complicated in the
case of a late-type galaxy's stellar population, which will be composed of many
different stars of varying ages and metallicities.  A composite 
CSP requires some knowledge of the SFH of the galaxy (to
assign an SSP at each timestep) and a chemical enrichment model (to assign a
metallicity to each generation, see Lee \etal 2023).  

Two advances in recent years have dramatically improved our ability to deduce stellar
mass from photometry for galaxies.  The first is the increased sophistication in the
suite of stellar isochrones used to produce stellar population models that allows
inspection of the effects of exotic components, such as horizontal branch and blue
straggler stars, as well as more detailed understanding of the effects of dust and
the initial mass function (see Conroy \& Gunn 2010, hereafter CG10). In addition, there
is an awareness that detailed spectral energy distribution (SED) fitting is not
critical to deducing $\Upsilon_*$ from population models, so that broad-band
photometry can be adequate (Gallazzi \& Bell 2009) as long as sufficient wavelength
coverage is obtained. With this technique, one determines $\Upsilon_*$ through
various mass-to-light versus color relations as well as from detailed spectral
indices, a procedure that is still laced with complications (see McGaugh \& Schombert
2014).  One clear result from these photometric studies is that the dynamic range in
$\Upsilon_*$ becomes narrower with increasing wavelength; however, beyond 4$\mu$m the
effects of hot dust decouple the photometry from stellar atmospheric luminosities and
limits the ability to deduce $\Upsilon_*$.

Our study follows the procedures described in Schombert, McGaugh \& Lelli (2022;
hereafter, SML), which builds on the techniques focused on the near-IR filters
developed in Schombert \& McGaugh (2014).  For the reasons stated above, and outlined
in Schombert \& McGaugh (2014), the optimal wavelength to deduce $\Upsilon_*$ is
between 1 to 4$\mu$m, i.e. photometry from the near-IR filters of $JHK$ plus the
regions covered by WISE W1 and {\it Spitzer} (IRAC 3.6$\mu$m).  Other studies (e.g.
Taylor \etal 2011) have focused on full SED fitting from the UV to the far-red.  This
has the advantage of being able to follow recent star formation plus a broader
section of the SFH of a galaxy (with various optical filters being sensitive to
varying and shorter timescales), but has the disadvantage of working with fluxes at
wavelengths where the dust extinction is large and the $\Upsilon_*$ has a large
dynamic range such that small errors in the photometry produce large uncertainties in
$\Upsilon_*$.  Taylor \etal conclude that the use of near-IR photometry increases the
uncertainty in the $\Upsilon_*$ value from full SED fits.  SML reached a similar
conclusion, but found fault was not in the near-IR fluxes but rather the stellar
population models in the near-IR (see also McGaugh \& Schombert 2014).  The model
uncertainties developed from inadequate treatment of the AGB population and,
therefore, SML used an empirical scheme to link the stellar population models to the
$K$, W1 and IRAC 3.6$\mu$m filters.

Our stellar population models is based on two main assumed functions, the SFH and the
chemical evolution as a function of time.  Our star formation history scenarios (SFR
as a function of age) fall into two classes:  1) the exponentially declining SF
models outlined in Speagle \etal (2014) for galaxies with stellar masses greater than
$10^{10} M_{\sun}$ and 2) a slowly declining, nearly constant SF model to match the
characteristics of the main sequence for galaxies with less than $10^{10} M_{\sun}$.
All the models assume an initial epoch of star formation of 12 Gyrs ago ($z_f$
between 8 and 12) followed by a rapid burst which declines or levels off by 4 Gyrs
(see Fig. 1 in SML).  The age-metallicity relation follows the scenario proposed by
Prantzos (2009) starting with a pre-enriched population of [Fe/H]=$-1.5$ followed by
a rapid rise to 80\% of the final metallicity (given by the mass-metallicity
relation).  Using these SFH and metallicity prescriptions, one can combine various
SSP's to reproduce both the main sequence (SFR versus galaxy stellar mass) and the
present-day two-color diagrams, the observational constraints on the modeling (see
SML).

The uncertainty in these SFH models can be compared to the range in SFR from the main
sequence (McGaugh \etal 2017) and the scatter in galaxy color from the ridgelines in
the two-color diagrams.  Numerical experiments changing the SFR and metallicity
growth by 30\% or varying epochs of quasi-steady bursts, can recover the scatter in
the main sequence and two-color diagrams (both measures of the present-day state of a
galaxy's stellar population).  These effects are important in two realms, the optical
portion of the spectrum, which is sensitive to the short-lived components of a stellar
population, and for low mass dwarf galaxies, where there is already evidence that
their SFH's are sporadic (see HST CMD's, Weisz \etal 2011).  However, it is important
to note that the {\it integrated} properties of galaxies is surprisingly narrow and
coherent.  Relationships, like the main sequence and the mass-metallicity relation,
indicate that, even if the star formation path in late-type galaxies is erratic,
ultimately they end up with a constant percentage of gas converted into stars at a
relatively steadily rate.  This produces a final stellar population of similar
distribution in age and metallicity in proportion to the total stellar mass.  This is
another reason for focusing on near-IR fluxes to determine stellar mass as those
wavelengths are most sensitive to the stable, long-lived portions of a stellar population,
minimally effected by dust extinction from recent SF and have the smallest percentage
change in $\Upsilon_*$ in models with varying SFH.

The most important component for the near-IR stellar populations is the treatment of
thermally pulsating AGB stars.  These are stars in the very
late stages of their evolution powered by a helium burning shell that is highly
unstable.  AGB stars have high initial masses ($M > 5 M_{\sun}$) and are intermediate
in age ($\tau > 10^8$ yrs).  While the Bruzual \& Charlot (2003; hereafter, BC03)
codes (and their extension, see Bruzual 2009) include AGBs as part of their
evolutionary sequence, comparison with other codes (e.g. Maraston 2005,
Gonzalez-Perez \etal 2014) finds discrepancies in the amount of luminosity from
this short-lived population.  The history of AGB treatment in SED codes is outlined
in CG10 and the work of Salaris \etal (2014) outlines the difficulty in AGB modeling,
primarily due to 1) variations of brightness and color on timescales of a few
thousand years, 2) their cool temperatures that have a dominant effect mostly in the
near-IR (where spectral libraries are less complete) and 3) the formation of dusty
envelope.  All these factors significantly complicated the SED calculations.  This is
most certainly the origin of the disagreement between population characteristics
deduced by Taylor \etal for optical versus near-IR colors.  

\begin{figure}
\centering
\includegraphics[scale=0.85,angle=0]{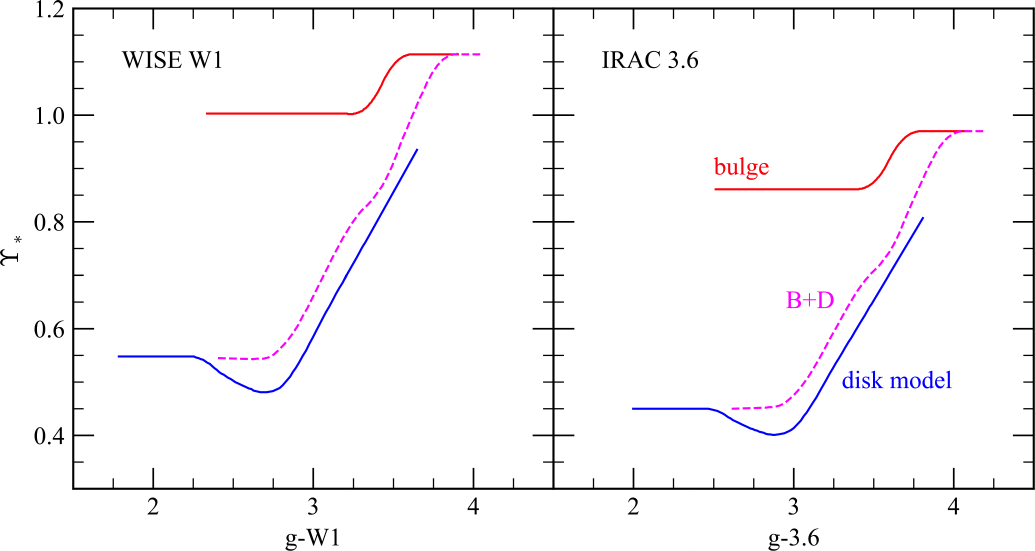}
\caption{\small Updated $\Upsilon_*$ models from Schombert, McGaugh \& Lelli (2022).
The three models shown (discussed in the text) represent pure bulge and disk star
formation histories, which are then combined to produce a bulge+disk (B+D) scenario
guided by the main sequence relation (Speagle \etal 2014) and a chemical enrichment
model.  Standardized to the $g$-W1 or $g$-[3.6] color, at the 3.5$\mu$m wavelength
the difference between $\Upsilon_*$ is only larger by 10\% from IRAC 3.6$\mu$m to WISE W1.
}
\label{ml}
\end{figure}

Due to these complications, we have adopted a more empirical approach to the AGB
component.  As noted in Schombert (2016) and Schombert \& McGaugh (2014), all the
recent stellar population models were poor matches to the near-IR colors of
populations younger than a few Gyr.  To best determine which stellar population model
matches our empirical AGB prescription, we compared extensions of the BC03 and
CG10 models to the near-IR colors of LMC, SMC and MW star clusters. As metallicity
typically decreases for older star clusters, it is problematic to compare a single
metallicity SSP track to the colors versus age diagram. However, a solar metallicity
model accurately captures the young cluster colors and conforms to a majority of the
expected metallicity of stars in normal galaxies. There is very little difference
between the solar BC03 and CG10 tracks, although adding a standard AGB dust model
(Villaume \etal 2015) compares more favourably with the redder young cluster colors.
We found the variation in metallicity is only important for very young and very old
clusters (see Fig. 6 in Schombert, McGaugh \& Lelli 2019).  As the timescale for
these types of stars is short, their effects are most important for the last
generation of stars in a galaxy, where the metallicity has reached its highest value
by galaxy mass.  This confines the range of AGB tracks required to reproduce near-IR
fluxes.

Using these 116 star clusters, we tabulate a set of $V-K$ colors for a range of age
and metallicities which we apply to our stellar population algorithm that builds a
composite population with a selected SFH and chemical enrichment scenario (see SML).
For IRAC 3.6$\mu$m flux, we built empirical $K$-[3.6] versus $V-K$ or $g$-$K$ color relations
using S$^4$G (Sheth \etal 2010) and SPARC samples.  We apply these relationships to
the $K$ population models to calculate $\Upsilon_*$ at 3.6$\mu$m.  A reality check is
provided by the fact the stellar population models correctly predict the various
optical to 3.6$\mu$m colors over a range of star formation and chemical evolution
scenarios.

Lastly, we perform the same operations for this paper on the W1 filter set, using the
color relations outlined in Paper I.  Our final $\Upsilon_*$ diagrams are shown in
Fig. \ref{ml}.  Differing from our previous papers, we present the $\Upsilon_*$ color
relations using SDSS $g$ and W1 rather than $V-$[3.6].  The difference from $V-$[3.6]
to $g-$[3.6] is small, and either color or morphological type is equally suitable for
deducing $\Upsilon_*$ (see Fig. 2 from SML).  The trends in Fig. \ref{ml} are similar
to previous color-$\Upsilon_*$ studies, a sharp difference between star-forming and
passive galaxy populations (disk versus bulge) and a steady, rising $\Upsilon_*$ with
redder color.  The short decrease in $\Upsilon_*$ for blue colors is a distinct
feature of the medium-aged AGB stars dominating the near-IR fluxes in high SFR
regions, a similar feature is seen in the optical from high-mass OB versus A stars.
While we have displayed our declining SF models, constant or enhanced SFR will
significantly altered the blue end of the color-$\Upsilon_*$ relation by $\pm$0.08
dex (see Fig. 2 from SML) even in the near-IR.

Three $\Upsilon_*$ scenarios are shown in Fig. \ref{ml}; a pure bulge model, a pure
disk model and a hybrid bulge+disk model.  Both the bulge and disk models are based
on a simplistic interpretation of either a uniform, old stellar population (bulge)
or a young, star-forming (i.e., dominated by massive stars) stellar population (disk).
Variation in color for the bulge scenario is due to increasing the mean metallicity
of the stars from [Fe/H] = $-1.5$ to super-solar.  The jump near $g-W1=3.3$ is due to
a sudden drop in the luminosity of AGB stars at near solar metallicities, confirmed
by the color of high-mass ellipticals.  For the star-forming disk scenario, the rise
in $\Upsilon_*$ with color is a combination of stronger star formation (Speagle \etal 2014)
with high-mass galaxies (i.e., more SF in the past and redder present-day colors) and
higher final metallicities (Cresci \etal 2019).  These models, for a range of
metallicities and SFH's, can be found at the SPARC website ({\tt http://astroweb.cwru.edu/SPARC}).

The bulge and disk models were produced in order to be applied to particular portions
of a galaxy's luminosity profile to give a combined spatial and luminosity weight to
a deduced value of $\Upsilon_*$.  However, many galaxy surveys have limited spatial
resolution (particularly at high redshift), and these datasets frequently only offer
a total luminosity value.  In order to more accurately produce a total stellar mass
from a total luminosity, we used the fact that the color and morphology of galaxies
are strongly correlated, and that relationship, in part, reflects the correlation of
B/D ratio with color.  A hybrid bulge+disk (B+D) model is produced by combining the
stellar populations from the pure bulge and disk models as a function of the B/D
ratio, plotted against color.  Unsurprisingly, increasing B/D also reflects into
redder colors (Graham \& Worthey 2008) and, to first order, one could substitute
morphological type in the place of $g-W1$ color and produce roughly the same
$\Upsilon_*$ (to $\pm$10\%).  The B+D models are shown in Fig. \ref{ml} and notice
the obvious behavior on the blue side where late-type galaxies with small or
non-existent bulges converge on the pure disk model.  Likewise, the early-type
galaxies with high B/D ratios merge into the pure bulge scenario and the presence of
even a small bulge in Sc type galaxies allows a monotonic connection from early to
late-type systems.

\section{Comparison of WISE Conversion to Stellar Mass}

Central to our technique is the use of color as a proxy for the metallicity and mean
stellar age to deduce $\Upsilon_*$.  As outlined in SML (see Fig. 9), the
mass-metallicity and mass-SFR relationships are sufficiently narrow to limit the range
in the characteristics of the underlying stellar population and predict a fairly
well-defined $\Upsilon_*$ versus color relation.  This has been demonstrated by
several authors in the optical (Bell \& Jong 2001; Taylor \etal 2011) and in the
near-IR (Eskew \etal 2012; Meidt \etal 2014) but is highly dependent on the smooth,
coherent evolution of galaxy stellar populations.  And there is every expectation
that the uncertainty in the models grow with decreasing galaxy mass as their star
formation history becomes more erratic (Weisz \etal 2011).

Our $\Upsilon_*$ models are based on optical to near-IR colors, confined by our
optical versus near-IR two color diagrams.  However, an alternative scheme to use
$W1-W2$ colors produces similar results.  The WISE study from Cluver \etal (2014)
showed that $W1-W2$ is well correlated with log $\Upsilon_*$ deduced from GAMA stellar
masses determined by SED fits from Taylor \etal (2011).  While there were two
distinct populations representing passive and star-forming galaxies, a linear trend
was evident and maximum likelihood fit of log $\Upsilon_* = -1.96(W1-W2)-0.03$ was
obtained (see their Fig. 6).

\begin{figure}
\centering
\includegraphics[scale=0.85,angle=0]{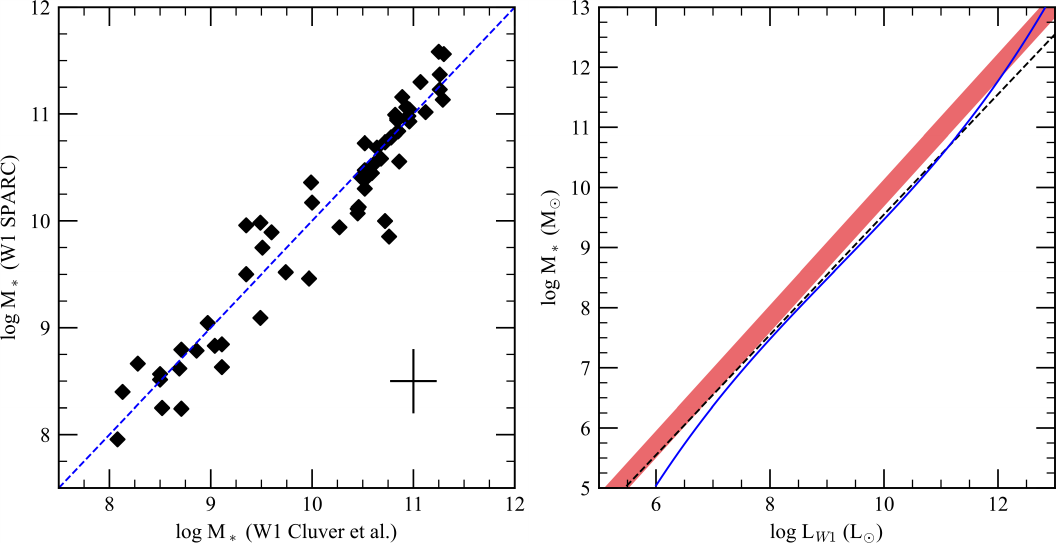}
\caption{\small A comparison of stellar masses deduced by $g-W1$ colors (SPARC)
versus $W1-W2$ color relation (Cluver \etal 2014).  The WISE sample was selected for
only those galaxies in the SPARC sample with $W2$ fluxes with S/N greater than 20.
Typical errors in stellar mass are shown and the RMS for the comparison of 58
galaxies is 0.29 in log.  The right panel displays the log $L_{W1}$ versus stellar
mass relationship outlined in Jarrett \etal (2023) (magenta line, a polynomial fit to
the GAMA masses).  Our WISE models are shown as a red band of $\pm$0.2 dex error
width based on a $g-W1$ color-magnitude relation to obtain $L_{W1}$.  The dashed line
indicates a constant $\Upsilon_*$ = 0.35.
}
\label{jarrett}
\end{figure}

We can use the WISE colors for the SPARC/WISE sample to compare the stellar masses
inferred from our optical to near-IR color technique to Cluver \etal WISE color
technique.  The results are shown in Fig. \ref{jarrett}, where we used a similar flux
cutoff in S/N at W2 as Cluver \etal and used the mean errors in our fits to determine
a mean error bar.  As can be seen in Fig. \ref{jarrett}, the correspondence is
excellent with a limiting RMS between the sample of 58 galaxies of 0.29 in log.  This
may represent the limit on stellar mass calculations using different models and
assumptions, but reinforces our earlier conclusion that optical or near-IR determined
stellar masses are in agreement for low redshifts.

A different relationship between W1 fluxes and stellar mass is found by Jarrett \etal
(2023).  In that study an improved version of the GAMA stellar masses is used
(GAMA-G23, Robotham \etal 2020) and new WISE total magnitudes from Jarrett \etal
(2019).  They follow a series of interlocking techniques to determine stellar mass
such as a straight-forward log $L_{W1}$ versus log $M_*$ relationship (a 3rd order
polynomial, but very close to a linear relationship), a revised $W1-W2$ color scheme
and a new $W1-W3$ color relation.  The $W1-W2$ colors have the same range and mean as
the SPARC sample, so our photometry is in agreement.  However, our resulting stellar
masses are about 0.2 dex higher than their total flux relation.  This can be seen in
the right panel of Fig. \ref{jarrett}, which shows the log $L_{W1}$ relationship
from Jarrett \etal fit in comparison to our $\Upsilon_*$ models where
we use the $g-W1$ color-magnitude relation as a proxy for absolute luminosity,
$L_{W1}$.  A constant $\Upsilon_*$ value of 0.35 is shown to demonstrate the
difference between our higher ($\Upsilon_* > 0.4$) values at low luminosities.

These new WISE stellar mass estimates from Jarrett \etal are difficult to
reconcile with $\Upsilon_*$ values deduced for $K$ and IRAC 3.6$\mu$m.  For example, the
range in $K-$[3.6] colors is between 0.1 and 0.4 (SML) which, when converted into solar
luminosities, corresponds to a mean luminosity at $K$ that is 1.5 times higher than
[3.6] (these also correspond to the observed SED's, Brown \etal 2014).  Our previous
$\Upsilon_*$ models have a value of 0.45 at [3.6] for a late-type galaxy with a $K-$[3.6]
color of 0.3.  The corresponding $\Upsilon_*$ for $K$ is 0.6 which produces identical
stellar masses from $K$ and 3.6$\mu$m luminosities (given the absolute magnitude of the
Sun, Willmer 2018).  Since the typical $W1-$[3.6] color for a late-type galaxies is
0.15, corresponding to a W1 flux in solar luminosities that is that 12\% fainter than
at 3.6$\mu$m (they have identical solar absolute magnitudes).  Therefore, the expected
$\Upsilon_*$ value should be near 0.5 for W1, in agreement with our models, but
slightly higher than the value of 0.2 predicted by Jarrett \etal (2023).  This may
signal a systematic difference in our W1 fluxes (unlikely) or a systematic
underestimate of near-IR $\Upsilon_*$ values from optically determined SED fits.  In
either case, this would also result in a sharp disconnect between stellar masses
calculated from {\it Spitzer} luminosities compared to WISE, as will be explored in
the next section.

\begin{figure}
\centering
\includegraphics[scale=0.85,angle=0]{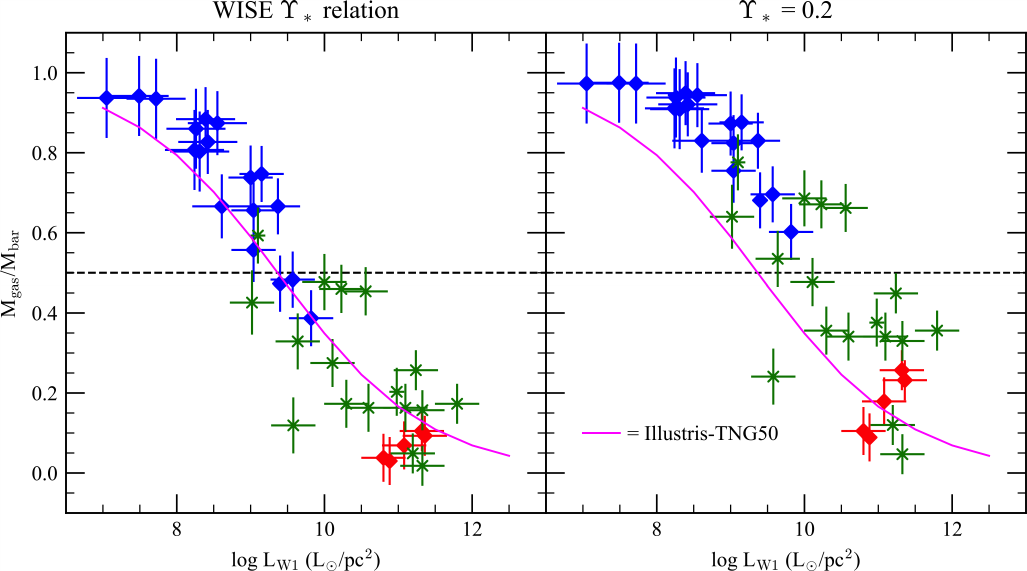}
\caption{\small A comparison of W1 luminosity versus gas fraction,
M$_{gas}$/M$_{bar}$, for the WISE SPARC sample.  The left panel displays gas
fractions using the color relation presented in this paper, while the right panel
uses a constant value of 0.2 proposed by Jarrett \etal (2023).  Hubble types are
indicated such that early-types spirals are red symbols, Sc's are green and late-type dwarfs
are blue.  The prediction of gas fraction as a function of luminosity from the
Illustris-TNG50 galaxy formation simulations (Korsage \etal 2023) is shown as a
magenta line.
}
\label{gas_frac_ml}
\end{figure}

Another avenue is to consider the change in the gas fraction produced by low
$\Upsilon_*$ values.  Following the analysis of Lelli \etal (2016), the gas fraction
($M_g/M_{bar}$) is shown in Fig \ref{gas_frac_ml} as a function of galaxy W1
luminosity in solar units.  Using our WISE $\Upsilon_*$ relationship based on galaxy
color produces the data in the left panel.  The gas fraction anticorrelates with
luminosity in a predictable manner with Hubble type.  One expects that early-type
spirals, being gas-poor, bottom out at high luminosities, likewise gas-rich,
late-type galaxies flatten out at low luminosities.  An expected S-shaped change in
gas fraction from early to late-types is seen.  This closely matches the predictions
from Illustris-TNG50 galaxy formation simulations (Korsga \etal 2023).

The lower $\Upsilon_*$ proposed by the GAMA-G23 stellar masses has the effect of
raising the gas fractions for all galaxy types (although significantly for late-type
dwarfs).  Rather than an S-shaped progression from early to late-types in terms of
gas fraction, the resulting new relationship has more of a box-like shape, a sharp
rise from Sa's to Sc's.  In addition, over 1/3 of SPARC galaxies would have gas
fractions above 0.5 (gas dominated disks, dashed line, Courteau 1996) for $\Upsilon_*
= 0.2$.  This is very strange for spiral galaxies because density wave theory (Lin \&
Shu 1964) suggests that well-developed spiral patterns cannot be supported in heavily
gas-dominated disks.  In summary, low $\Upsilon_*$ values are difficult to reconcile
with known gas and kinematic relationships for gas-rich galaxies.

\section{Comparison of WISE to {\it Spitzer} Total Stellar Masses}

While the same population models are used to deduce the stellar masses through the W1
or IRAC 3.6$\mu$m filters, the calibrations depend on the accuracy of the photometry and
interpretation of the color-color relations from Paper I.  As discussed in SML, three
different models can be used depending on the quality of the photometry.  If the
galaxy in question is fully resolved, then a bulge model can be used to assign a
bulge stellar mass and a disk model can be applied to the star-forming region to
assign a disk mass.  The two components can be summed for a total stellar mass.  If a
full surface brightness profile is lacking, the B/D ratio can be estimated from the
galaxy's color or morphological type (Graham \& Worthey 2008) and a combined model
(labeled B+D in Fig. \ref{ml}) can be applied to the total luminosity of the galaxy.
If using the total color, the model uses the B/D to color relationship from
SML to deduce a B/D luminosity ratio and sum the
resulting stellar from the arithmetic sum of the two components.  As will be shown in
the next paper of this series, this can have a significant impact on the deduced
stellar masses of early-type spirals with their large B/D ratios.

\begin{figure}
\centering
\includegraphics[scale=0.85,angle=0]{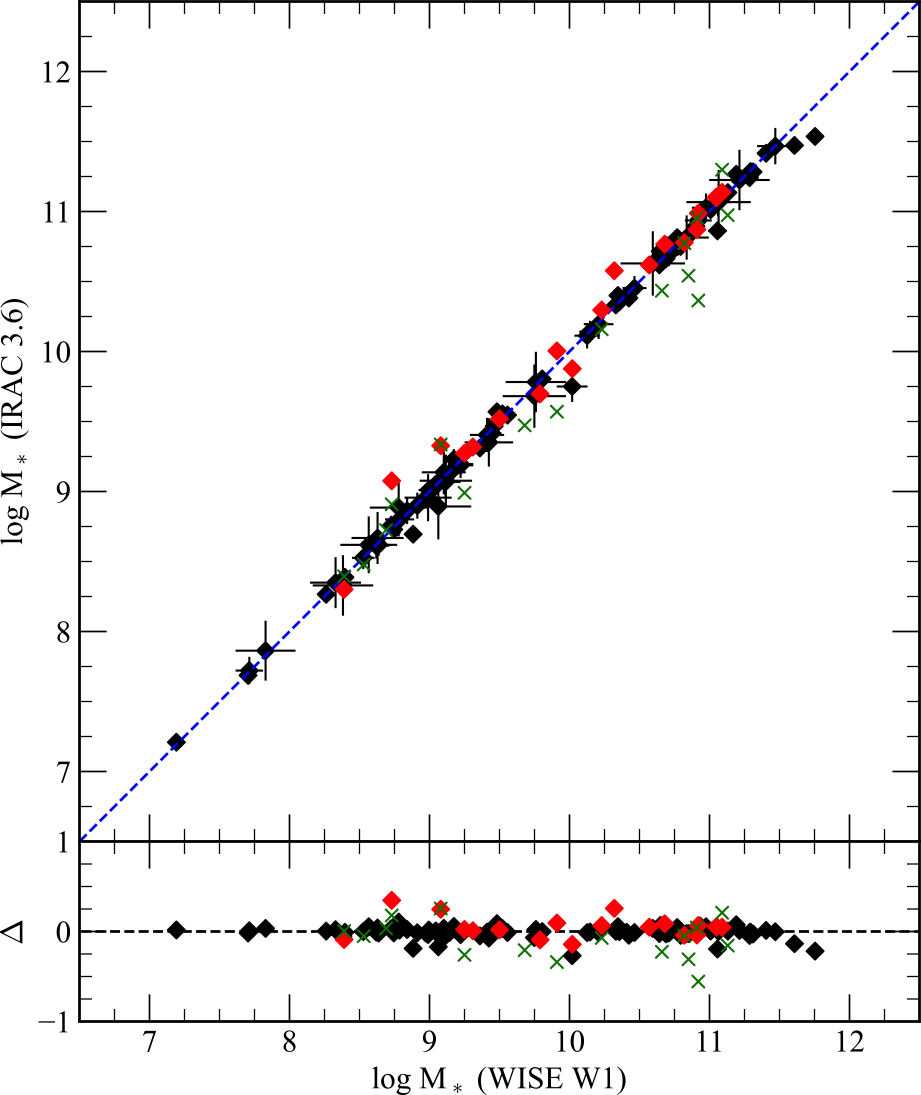}
\caption{\small Using the models from Fig. \ref{ml}, the IRAC 3.6$\mu$m and WISE W1
total luminosities are converted into total stellar masses in an independent fashion.
Due to the typically redder and narrow range of $W1-$[3.6] colors plus only slightly higher $\Upsilon_*$ values, it is
unsurprising to find the relationship between stellar mass at W1 and 3.6$\mu$m has a slope
of one.  The error from a slope of unity is $\pm$0.01 and the scatter is completely explained by
photometric errors. The red and green symbols are 20 and 16 galaxies in common from
the S$^4$G and Dustpedia surveys.  The residual diagram at the bottom shows no
systematics over the range of stellar masses with a mean offset of 0.02$\pm$0.18 dex.
}
\label{massc}
\end{figure}

As a check to our procedure, we use just the galaxy's $g-$[3.6] or $g-W1$ color to
assign a $\Upsilon_*$ for each galaxy in the SPARC sample with SDSS imaging.  We use
identical apertures for SDSS $g$, W1 and 3.6$\mu$m in order to measure the same
absolute fluxes (i.e., a metric color) and assign a comparable stellar mass.
The
resulting comparison stellar masses are shown in Fig. \ref{massc} and the deduced
stellar masses, colors and $\Upsilon_*$ values are listed in Table 1.  The
correspondence is excellent with the relationship of unity well within the
photometric errors.  The few outliers all have higher stellar masses in W1, probably
indicating that stellar contamination by foreground stars is still a serious concern
for W1 photometry and can overestimate a galaxy's total luminosity without some
visual inspection of the region around the WISE images.

In addition to a comparison with our own {\it Spitzer} photometry, we can also
compare to stellar mass estimates in the literature.  We have selected the S$^4$G
stellar masses (Muñoz-Mateos \etal 2015) which used the Eskew \etal (2012) technique
for converting 3.6$\mu$m fluxes into solar masses.  There were 20 galaxies in common
with SPARC samples shown in red in Fig. \ref{massc}.  Also selected were 16 galaxies
in common with the Dustpedia survey (Clark \etal 2018) which used 42 filters from
GALEX to ALMA to determine the stellar mass from SED fitting.  Both samples are in
excellent agreement with our W1 stellar masses.  In particular, we note good
agreement with the S$^4$G sample which used a different formula for determining
$\Upsilon_*$, but over the same portion of the spectrum as our W1 data.  We also note
that Dustpedia uses the full range of optical to mid-IR fluxes and a superior SED
algorithm to determine $\Upsilon_*$, yet achieves the same values as our IR technique.

\section{Comparison of WISE to {\it Spitzer} Stellar Mass Densities}

Total and aperture fluxes are the focus of the vast majority of stellar mass literature, and
most of the stellar mass scaling relationships focus on integrated values.
However, recent discoveries on the baryon-dark matter connection involve
point-by-point comparisons with kinematics and baryon density as a function of
radius (see, for example, the radial acceleration relation, Lelli \etal 2017).  A comparison
between {\it Spitzer} and WISE stellar mass densities will provide a baseline for
future work in this arena.

The subtle differences between the WISE W1 filter (centered at 3.4$\mu$m) and the
IRAC 3.6$\mu$m filter (centered at 3.55$\mu$m) are outlined in Paper I.  Briefly, the two
filters overlap for 87\% of their flux with W1 having a longer blue side and IRAC
3.6$\mu$m
covering slightly more wavelength on the redside (see Fig. 1, Paper I).  Both
filters cover a known PAH feature at 3.3$\mu$m, otherwise there are no strong
emission line features in this part of a normal galaxy's SED.

On the Vega magnitude system, zero $W1-$[3.6] colors are represented by a steep SED
that falls off to the red.  Galaxies with red, old stellar populations will have
$W1-$[3.6] colors near zero.  Galaxies with notable star formation will have slightly
more flux redward of the 3.3$\mu$m PAH feature due to young AGB stars and
contamination from hot dust redward of 4$\mu$m.  This produces slightly more flux in
the IRAC 3.6$\mu$m filter than W1, resulting in slightly redder $W1-$[3.6] colors.  This also
produces a reverse expectation from galaxy colors, blue $W1-$[3.6] colors signal old
populations like ellipticals or spiral bulges and red $W1-$[3.6] colors signal star
formation effects.

The SPARC sample is selected for rotational kinematics and covers all types of disk
and dwarf galaxies with Hubble type Sa to Irr.  As shown in Paper I, this produced a
$W1-$[3.6] color that was primarily on the redside of zero (between 0.15 and 0.25), where
the mean elliptical color is 0.1, see Fig. 8 in Paper I).  Low surface brightness
(LSB) galaxies displayed slightly redder $W1-$[3.6] colors than bright spirals.
Comparison with stellar population models (Schombert \& McGaugh 2014) indicates this
is a combined SFR and metallicity effect with lower mean metallicity associated with low star
formation and low stellar mass systems.  Given the above trends in color, surface
brightness and mass, we anticipate weak $W1-$[3.6] color gradients in spiral disks but,
in general, the W1 flux should track the IRAC 3.6$\mu$m flux.

All 175 galaxies in the original SPARC sample have WISE W1 images from the ALLWISE
archive (Duey \etal 2023).  However, a number of these galaxies have nearby stars or
foreground galaxies which make quality reduction of the WISE images impractical.  Of
the original SPARC sample, we isolated 111 galaxies with good IRAC and W1 images.  In
addition, 81 of that subsample also have SDSS DR16 $ugri$ images for optical
comparison.  

\begin{figure}
\centering
\includegraphics[scale=0.85,angle=0]{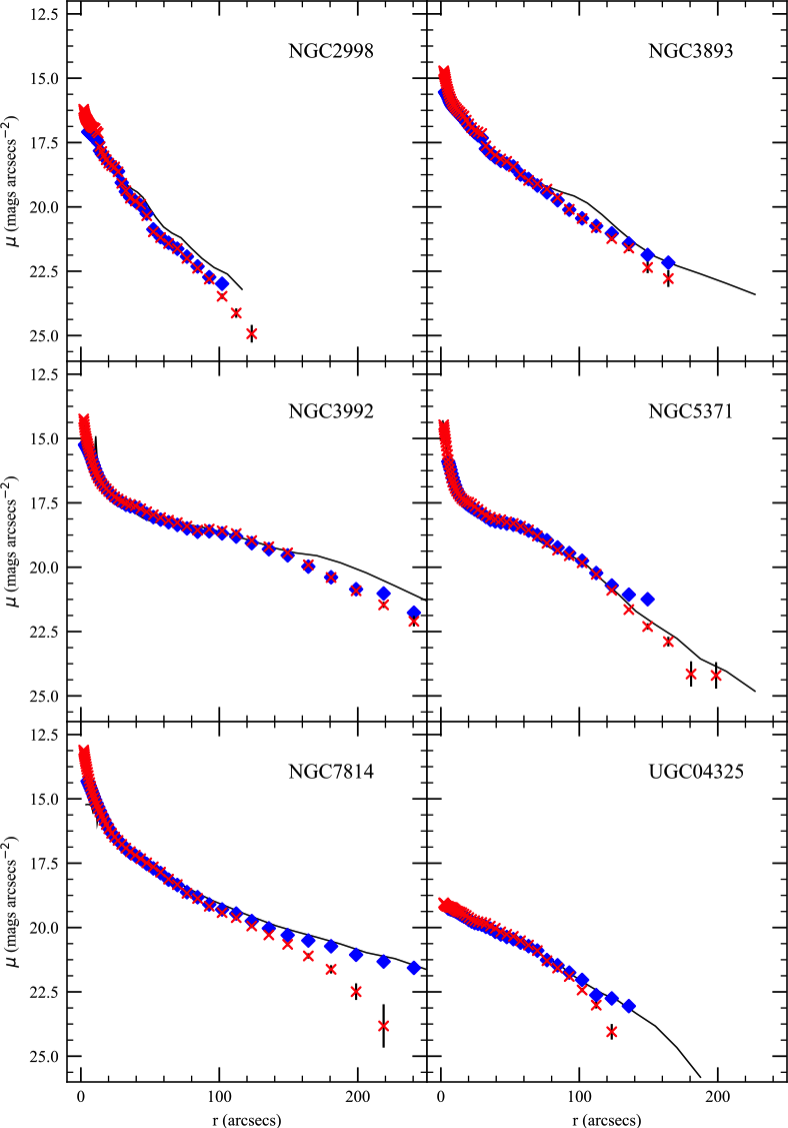}
\caption{\small The WISE W1 surface brightness profiles (blue symbols) are compared
to {\it Spitzer} IRAC 3.6$\mu$m (red symbols) and SDSS $i$ (black lines) profiles for six
SPARC galaxies of varying mass, morphological type and scale length.  The W1 and
3.6$\mu$m
profiles are in excellent agreement with a small $W1-$[3.6] color offset.  All the same
stellar luminosity features detected in 3.6$\mu$m are seen in the W1 profiles.  The SDSS
$i$ profiles are offset by a $i-$[3.6]=2.5 color for comparison.   The correspondence
between the far-red and near-IR is also excellent, providing additional confidence
in the use
of optical imaging to follow profiles to smaller radii where WISE imaging lacks
resolution.
}
\label{sfb}
\end{figure}

A subset of eight galaxies with SDSS $i$, W1 and IRAC 3.6$\mu$m surface brightness profiles
are shown in Fig. \ref{sfb}.  These eight galaxies were selected to cover a range of
$W1-$[3.6]
color, size and profile shape.  Unsurprisingly, since their $W1-$[3.6] colors are near
zero, all the W1 and IRAC 3.6$\mu$m profiles have nearly the same surface brightnesses (for
clarity we interpolated the W1 radii to match the IRAC 3.6$\mu$m radii).  The exact same
luminosity features seen in IRAC 3.6$\mu$m profiles are visible in WISE W1 profiles.
The W1 profiles were truncated when the surface brightness errors exceeded 0.25 mags
arcsecs$^{-2}$, which was due to a combination of lower S/N and greater uncertainty
in the W1 sky value (sky uncertainty dominates photometry of galaxies over 1 arcmin
in size, see Schombert \& McGaugh 2014).

The black lines in Fig. \ref{sfb} are the SDSS $i$ surface brightness profiles
shifted by a mean color of $i-$[3.6] = 2.5.  The correspondence is excellent across all
three filters with some expected deviation at large radii due to weak $i$ band color
gradients.  This gives some confidence to using far-red photometry to follow the
stellar mass in high redshift samples.  We note that the color-$\Upsilon_*$ relation
for SDSS $i$ has a factor of five larger dynamic range compared to W1 (see Fig. 3 SML) with
a proportionate amount of error in $\Upsilon_*$ for photometric errors in color.  A
optimal course of analysis would be to use higher resolution optical imaging for
local stellar mass determination tied to stellar mass values deduced from coarser,
aperture near-IR values.

\begin{figure}
\centering
\includegraphics[scale=0.85,angle=0]{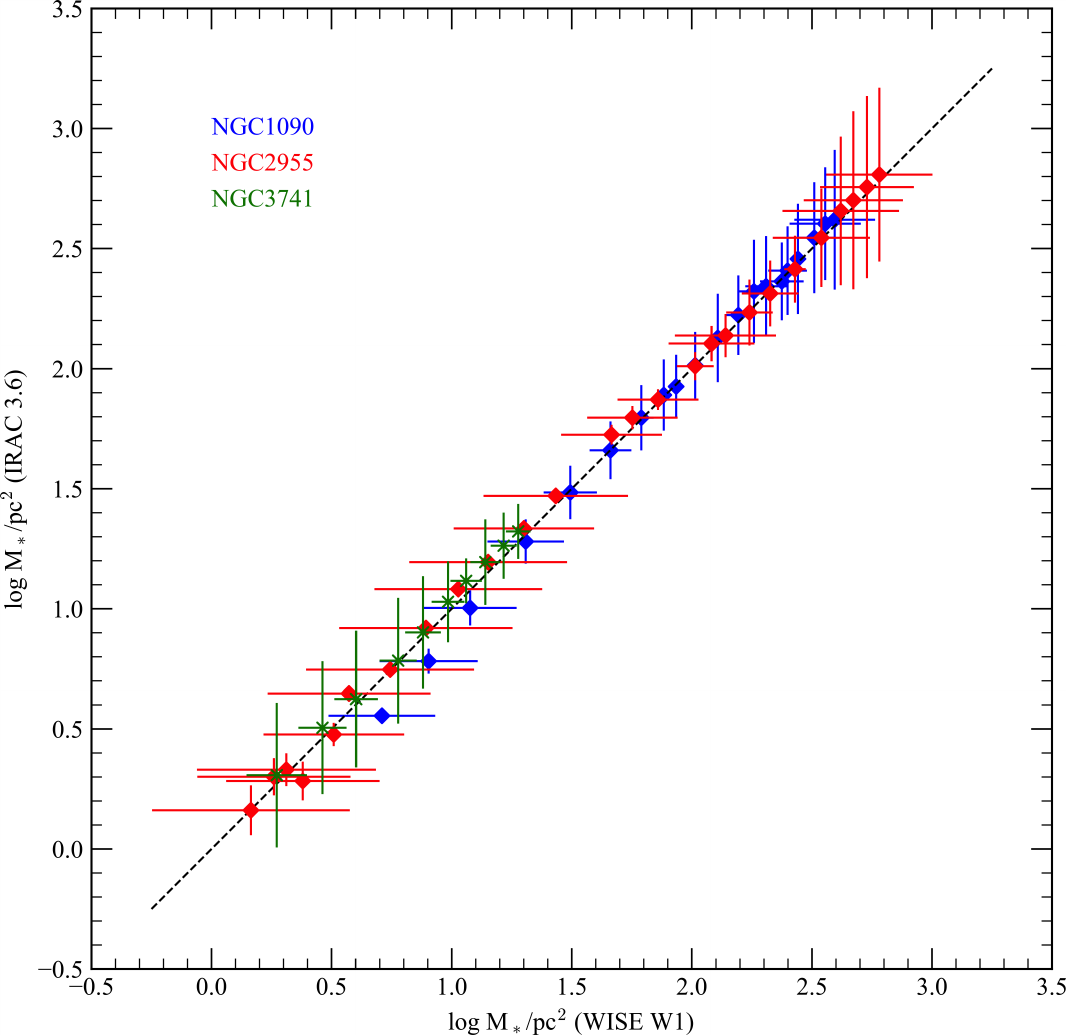}
\caption{\small The comparison of three SPARC galaxies mass density profiles using
WISE W1 surface brightness profiles versus {\it Spitzer} IRAC 3.6$\mu$m profiles.
}
\label{lum_profile}
\end{figure}

The conversion of surface brightness profiles into stellar mass density profiles is
similar to the procedure for deducing total stellar masses (without the need for a
galaxy distance as surface brightness is distance independent).  As shown in Fig.
\ref{lum_profile}, three galaxies from the SPARC sample with varying morphological
types and colors are selected for comparison of the IRAC 3.6$\mu$m versus WISE W1 mass
profiles.  The model $\Upsilon_*$ value for each isophote is determined from either
the $g-$[3.6] or $g-W1$ color and applied to each radius independent of the surface
brightness value in either W1 or 3.6$\mu$m.

Again, the correspondence is excellent, regardless of the total stellar mass, the
IRAC 3.6$\mu$m and WISE W1 values are in agreement within the observational errors
(determined from the uncertainties in the photometry of each isophote).  The WISE
data was ignored below the 5 arcsecs radius due to the poor spatial resolution of
WISE imaging.  Values below the PSF will always reproduce lower mass density values
as W1 flux will be distributed to larger radius by the PSF.  This results in
larger uncertainties at higher surface densities for the WISE imaging in Fig.
\ref{lum_profile}.  But the use of WISE W1 for determination of stellar mass
densities should be comparable to the mass models produced by {\it Spitzer} (Lelli
\etal 2016), with the caveat of decreased spatial resolution from the larger PSF for
WISE.

\section{WISE Tully-Fisher Relation}

The original form of the Tully-Fisher relation compared the HI line profile
half-width (W$_{50}$) versus stellar luminosity (typically for a filter in the red or
near-IR to minimize extinction effects, Aaronson \& Mould 1983).  The link between gravitating mass and
rotation is the underlying principle to the luminosity TF relation, where luminosity
is a proxy for stellar mass.  As the focus of many TF studies is rich clusters (see
CosmicFlow, Tully \etal 2008), then the spiral studied are predominately early-type
spirals.  Cluster spirals are typically gas-poor and high mass systems, ideally
positioned for distance scale analysis as a class of objects.

\begin{figure}
\centering
\includegraphics[scale=0.85,angle=0]{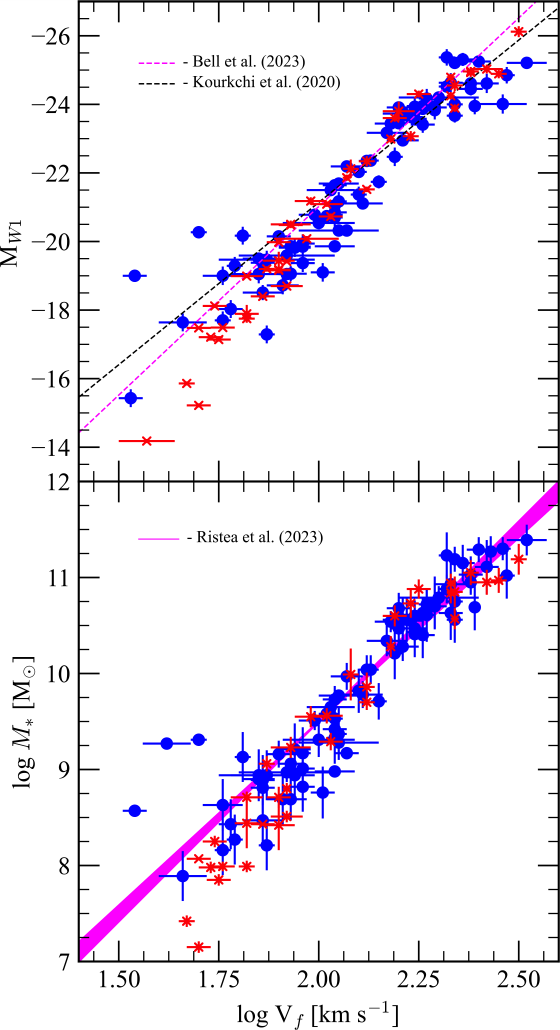}
\caption{\small The luminosity and stellar mass TF relations for the SPARC WISE
sample.  Blue symbols indicate galaxies with Hubble flow distances, red symbols are
galaxies with redshift-independent distances (typical Cepheids or TRGB estimates).
The luminosity TF is key as a distance indicator for high-mass cluster spirals 
from Kourkchi \etal (2020) and
Bell \etal (2023) are shown.  The bottom panel displays the same data converted into
stellar mass using the $\Upsilon$ values deduced as discussed in \S2.  For
comparison, the stellar mass TF from Ristea \etal (2023) is shown.  The range of fits
for stellar versus gas rotation curves is shown as the magenta band.
}
\label{lum_tf}
\end{figure}

The WISE luminosity TF relation is shown in Fig. \ref{lum_tf} for the SPARC sample
with WISE W1 photometry and accurate distances.  The sample is divided into two
sub-samples, those galaxies with redshift-independent distances and those galaxies
with Hubble distances from EDD (Extragalactic Distance Database, Tully \etal 2009).
The top panel displays the absolute W1
total magnitudes determined from curves of growth as outline in Paper I versus
$V_{f}$, the asymptotic velocity value from the SPARC HI rotation curve (Lelli
\etal 2019).  The
relationship is close to linear (in log space) with some indication of a change in
slope at $L_*$ ($M_{W1} = -22$).

A comparison to previous WISE TF fits is shown in Fig. \ref{lum_tf} from Kourkchi
\etal (2020, CosmicFlows3) and Bell \etal (2023).  Those studies focused on high
luminosity cluster spirals to map peculiar velocity in the local Universe and their
fits agree on the high luminosity end of our data.  Deviations at the low luminosity
end are more than likely due to differences in photometric assignment to a total
luminosity value for low surface brightness dwarfs, and differences between
rotation velocity from half-width velocities ($W_{50}$) and rotation curves (see
Lelli \etal 2019).

The bottom panel of Fig. \ref{lum_tf} displays the stellar mass TF relationship.  The
key difference in this version of the TF relation is that W1 luminosity has been
converted into stellar mass using the color prescriptions in \S2.  The stellar mass
TF relation differs from the luminosity TF in that low luminosity galaxies have
bluer colors, and corresponding low $\Upsilon$ values.  In addition, the high mass
spirals have significant bulges, which have higher $\Upsilon$ values and a
proportionally greater contribution to total stellar mass relative to their
luminosities.  This has the effect of raising the high mass galaxies with respect to
the low mass end, and improving the linearity of the entire correlation.

Also shown in the bottom panel of Fig. \ref{lum_tf} is the stellar mass fits from
Ristea \etal (2023).  For their study, stellar masses were assigned from MaNGA Galaxy
Survey (Bundy \etal 2015) WISE photometry using the spectral energy distribution
fitting Code Investigating GALaxy Evolution (CIGALE; Boquien et al. 2019).  Rotation
velocities were determined for the stellar component (from H$\alpha$ spectra) and the
gas component (from \hi spectra).  The two different velocity measurements resulting
in slightly different TF slopes as shown by the magenta area in Fig. \ref{lum_tf}.

As with the luminosity TF, the stellar mass TF agrees well on the high-mass end, but
over estimates the stellar masses (or under estimates the rotation velocity) on the
low-mass end.  As the SPARC data uses the flat portion of the rotation curve, an
under estimate is more likely.  The linearity of stellar mass TF also breaks at
approximately 10$^{10}$ $M_{\sun}$, at the same luminosity as the luminosity TF
deviates from linearity.  This is also the point where the typical gas fraction rises
about 20\% for spirals indicating a increasing importance of the gas mass of a galaxy
to its gravity and, thus, its rotation velocity.

\section{Summary}

Future work on deducing stellar masses from near-IR photometry will be
less dependent on pointed observations from {\it Spitzer} (due to the recent end of
mission) and more dependent on all-sky surveys at similar wavelengths, such as the WISE
mission. We present new stellar population models to produce a simple, and effective,
color to $\Upsilon_*$ at the WISE W1 wavelengths.  Our main results are:

\begin{enumerate}

\item We extend the stellar population models developed in Schombert, McGaugh \&
Lelli (2019) to cover the WISE filters.  This is a simple interpolation from our
previous models, empirically tied to the {\it Spitzer} IRAC 3.6$\mu$m colors and
calibrated to stellar cluster colors to correct for short-lived AGB populations
inadequately addressed by isochrone studies.

\item Our color-$\Upsilon_*$ models follow three scenarios that cover the range of
expected galaxy evolution.  They are a) a pure disk population, representing a slowly
declining SFR history whose final strength is set by the main sequence relation
(Speagle \etal 2014) and a metallicity set by the current gas-phase metallicity
(Cresci \etal 2019) with a chemical evolution scenario, b) a pure bulge population
composed of a passive, burst population that varies in color and $\Upsilon_*$ based
on the color-metallicity relation (see SML), and c) a combined bulge+disk model that
uses the well-known color to morphology relation to allow a user to estimate a mean
$\Upsilon_*$ from total color or morphological type.

\item These three scenarios are shown graphically in Fig. 1, where the most
notable difference from the {\it Spitzer} color-$\Upsilon_*$ relation is an expected
upward shift in $\Upsilon_*$ by approximately 0.1 dex.  The dynamic range is nearly
identical from IRAC 3.6$\mu$m to WISE W1, meaning that similar photometric errors will
reflect into similar errors in stellar mass for {\it Spitzer} versus WISE
observations.

\item Stellar masses deduced from WISE photometry for the SPARC sample are in
excellent agreement with our original stellar masses from {\it Spitzer} photometry
and other WISE stellar masses in the literature (e.g., Cluver \etal 2014).
This was not unexpected as the models are similar and the W1 to [3.6] colors are
consistent with the 0.1 dex increase in $\Upsilon_*$ predicted by the population
models.

\item Comparison with the S$^4$G survey stellar mass estimates (based on
3.6$\mu$m aperture fluxes and the Eskew \etal 2012 models) plus the Dustpedia stellar masses
(based on SED fits to 42 filters from GALEX to Planck) are also in excellent agreement
with our W1 fluxes and our own stellar population scenarios.  
Comparison to other WISE and {\it Spitzer} studies reinforce the importance of using
a galaxy color to confine the $\Upsilon_*$ models.

\end{enumerate}

There are several takeaways from our new WISE W1 models and comparison to existing
stellar mass estimates.  First, the dynamic range in the color-$\Upsilon_*$ relations
is still the smallest compared to optical filters, despite the
uncertainties in the AGB component, and produces the most stable $\Upsilon_*$ values for
the expected photometric errors.  This is true both for integrated values (total
galaxy stellar mass) and surface density profiles.  

Second, varying star formation histories are important for the low-mass, blue dwarf
galaxies (at the 0.1 dex level), but to reproduce present-day galaxy colors, there is
very little variation in $\Upsilon_*$ for high-mass, red spirals.  Accuracy for
early-type spirals is more dependent on separating the bulge and disk components
(with their differing color-$\Upsilon_*$ relations) using moderately high resolution
surface photometry.

Third, the similar values between different datasets (e.g, S$^4$G) and different
techniques (e.g., Dustpedia) indicate that choice of filter set or modeling are less
important than solid photometry data.  It is encouraging that $\Upsilon_*$ values
from the far-red to near-IR produce identical stellar mass values, as this region is
less influenced by distortion due to extinction.  This will also be critical in
linking stellar masses estimates at high redshift (i.e., JWST) using UV and optical
rest-frame photometry to present-day stellar mass deduced from near-IR fluxes.  The
full SED fitting techniques of Dustpedia allow for a stable connection between those
fluxes from deep space imaging and near-IR calibrators.

Lastly, the differences between the luminosity TF and stellar mass TF are shown in Fig.
\ref{lum_tf}.  The importance of the luminosity TF to distance scale work has been
demonstrated in numerous studies (see Tully 2023 for a review).  And the high-mass end
the smallest scatter since these systems have the smallest gas fractions and,
therefore, the stellar mass represents the total baryon mass which as the strongest
correlation with rotational velocity (see Lelli \etal 2019).  The conversion of luminosity to
stellar mass (also shown in Fig. \ref{lum_tf} does not significant improve the
linearity of the TF relation, again due to the varying contribution of missing gas
mass on the low luminosity end, but does improve the shape of the high-mass end with
the proper division of bulge versus disk $\Upsilon$ values and complete sum of the
bulge and disk components.

\section*{Acknowledgements}
Software for this project was developed under NASA's AIRS and ADAP Programs. This work
is based in part on observations made with the Spitzer Space Telescope, which is
operated by the Jet Propulsion Laboratory, California Institute of Technology under a
contract with NASA.  Support for this work was provided by NASA through an award
issued by JPL/Caltech. Other aspects of this work were supported in part by NASA ADAP
grant NNX11AF89G and NSF grant AST 0908370. As usual, this research has made use of
the NASA/IPAC Extragalactic Database (NED) which is operated by the Jet Propulsion
Laboratory, California Institute of Technology, under contract with the National
Aeronautics and Space Administration.

\begin{deluxetable}{lcccc}
\tablecolumns{6}
\small
\tablewidth{0pt}
\tablecaption{WISE Stellar Masses}
\tablehead{
\\
\colhead{Name} & \colhead{log $L_{W1}$ } & \colhead{log $M_{W1}$} & 
\colhead{$g-W1$} & \colhead{$\Upsilon_{W1}$} \\
 & \colhead{($L_{\odot}$)} & \colhead{($M_{\odot}$)} & & \\
}
\startdata
F568-1  & 9.67 & 9.41$\pm$0.06 & 2.50 & 0.55 \\
F568-3  & 9.79 & 9.52$\pm$0.03 & 2.73 & 0.54 \\
F568-V1 & 9.44 & 9.17$\pm$0.04 & 2.85 & 0.54 \\
F571-V1 & 9.05 & 8.79$\pm$0.05 & 2.77 & 0.54 \\
F574-1  & 9.67 & 9.48$\pm$0.03 & 3.16 & 0.65 \\
\enddata

\tablecomments{Only the first five galaxies are shown.  The rest are in the electronic version.}
\end{deluxetable}

\end{document}